\documentclass[pss]{wiley2sp} 
\usepackage{amsmath}
\usepackage{bm}              
\usepackage{w-greek}         

\begin{document}

\title{Emission spectrum of a quantum dot embedded in a nanocavity}

\titlerunning{Spectrum of a QD in a nanocavity}

\author{%
   Guillaume Tarel\textsuperscript{\textsf{\bfseries\Ast}},
   Vincenzo Savona}

\authorrunning{Guillaume Tarel et al.}

\mail{e-mail
  \textsf{guillaume.tarel@epfl.ch}, +41-21-6933423, Fax +41-21-6935419}

\institute{Institut de Th\'eorie des Phenom\`enes Physiques, Facult\'e des Sciences de Base, Ecole Polytechnique F\'ed\'erale de Lausanne
Station 3, CH-1015 Lausanne EPFL, Switzerland}

\received{XXXX, revised XXXX, accepted XXXX} 
\published{XXXX} 

\pacs{ } 

\abstract{%
%
%
%
\abstcol{We model the emission spectrum of a quantum dot embedded in a (e.g. photonic crystal) nanocavity, using a semi-classical approach to describe the matter-field interaction. We start from the simple model of a quantum dot as a two-level system, and recover the result expected from cavity quantum electrodynamics. Then, we study the influence of electron-acoustic-phonons interaction. We show that the surrounding semiconductor plays an essential role in the emission spectrum in strong coupling.
  }}

%
%

\titlefigure[width=0.3\textwidth]{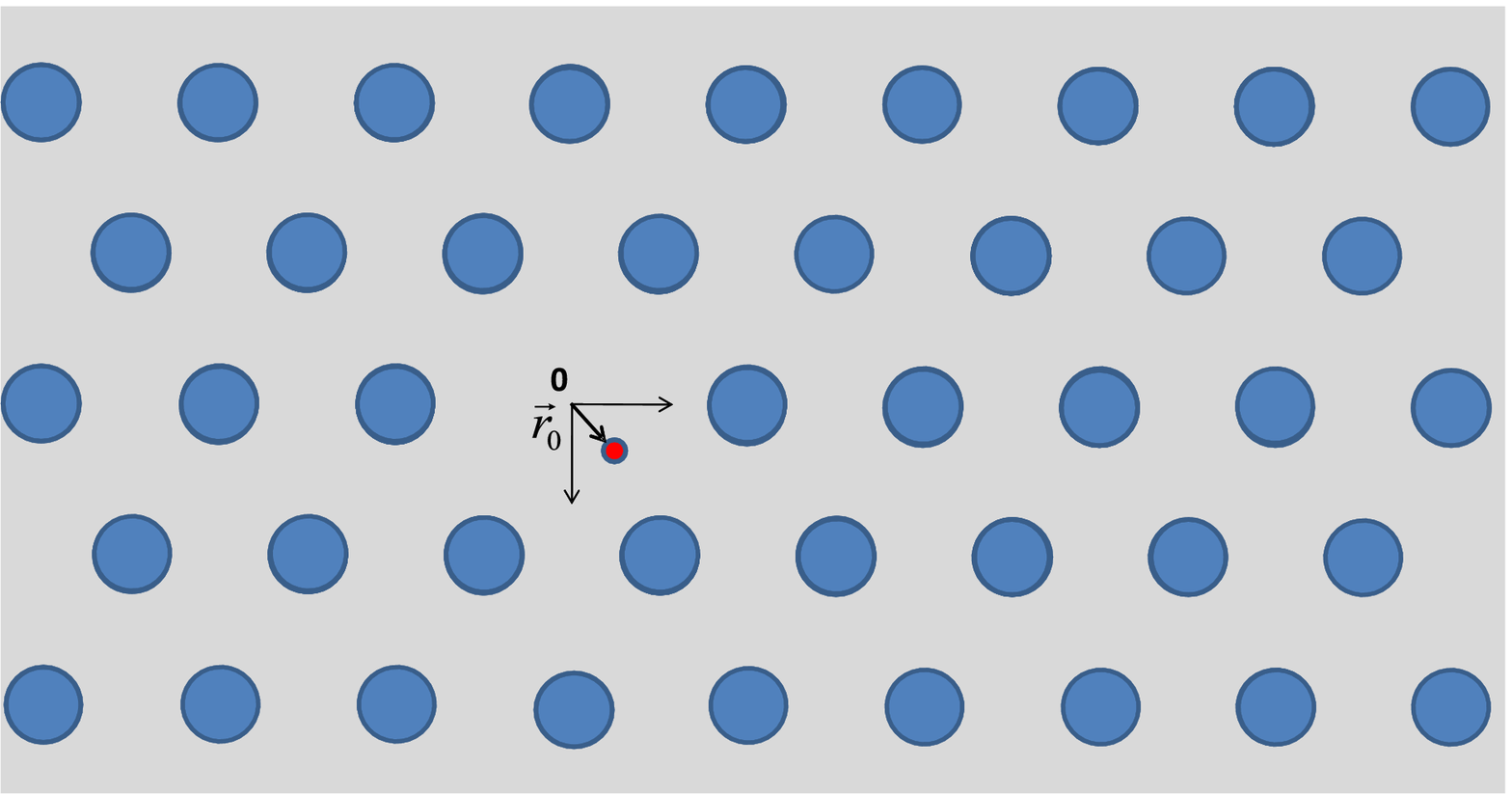}\label{fig1}
\titlefigurecaption{Schematic of the L1 cavity of the 2D triangular PHC with embedded quantum dot.}

\maketitle   

\section{Introduction}

Semiconductor quantum dots (QDs), displaying a quantized spectrum of energy levels, can be considered as a solid state equivalent of atoms. Recently, they allowed to draw a parallel between semiconductor physics and Cavity Quantum Electrodynamics (CQED, formerly being restricted to the domain of atomic physics). This was made possible by tailoring the light-matter interaction in resonant photonic structures such as micropillars and photonic crystal (PHC) nanocavities. Indeed, various cavity related effects such as Purcell effect and Rabi splitting have been experimentally observed \cite{Reithmaier2004}. Understanding how the semiconductor environment surrounding a QD affects CQED properties is of great interest, as it gives a measure of how the QD-cavity system departs from an ideal CQED system, thus assessing the feasibility of a QD implementation of quantum information technology.

In this work, we model the emission properties of a QD embedded in a nanocavity (e.g. a L1 defect in a PHC slab \cite{Andreani2004}). We extend a semi classical formalism originally developed to study radiative properties of QDs in vacuum and planar resonators \cite{Parascandolo2005}. The electromagnetic field is described in terms of Maxwell equations, and the dielectric structure of the nano-cavity is accounted for by means of the resolvent representation of the Green's function. A susceptibility tensor models the QD degrees of freedom. In the simple two-level assumption, the well known CQED result is recovered. By including the electron-phonon coupling, we discuss how acoustic phonon sidebands \cite{BorriPRL} in the QD affect the strong coupling.

\section{Theory}

The classical electric field ${\bm{\mathcal E}}\left({\bm{r}},\omega\right)$ obeys a usual Maxwell equation, where $\bm{r}$ is the position vector in 3-D, $\epsilon({\bm{r}})$ is the spatially dependent dielectric constant and $\hat{\bm{\chi}}_{QD}$ is the linear optical susceptibility tensor of the QD subsystem:
\begin{multline}\label{Max1}
\bm{\nabla}\wedge\bm{\nabla}\wedge
{\bm{\mathcal E}}\left({\bm{r}},\omega\right)-\frac{\omega^2}{c^2}
\bigg[\epsilon({\bm{r}}) {\bm{\mathcal E}}\left({\bm{r}},\omega\right)
\\
\left.+4\pi\int d{\bm{r}}^\prime
\hat{\bm{\chi}}_{QD}\left({\bm{r}},{\bm{r}}^\prime,\omega\right)
\cdot {\bm{\mathcal E}}\left({\bm{r}}^\prime,\omega\right)\right]=0\,.
\end{multline}

While $\epsilon({\bm{r}})$ is a constant in the case of vacuum, it describes in general the dielectric structure surrounding the QD. In the case of a PHC nanocavity -- that we consider here -- it describes the spatial profile of the dielectric constant that varies between a low and a high index material.

In order to obtain a Hermitian eigenvalue problem for the Maxwell equation, we define
\begin{equation}
 \bm{\mathcal Q}=\sqrt{\epsilon({\bm{r}})}\bm{\mathcal E}(\bm{r},t)\,,
\end{equation}
and
\begin{equation} \label{Max}
\bm{\Upsilon}=\frac{1}{\sqrt{\epsilon({\bm{r}})}}\bm{\nabla}\wedge \{ \bm{\nabla}\wedge \frac{1}{\sqrt{\epsilon({\bm{r}})} } \}\,.
\end{equation}

This leads to an equation for the hermitian linear differential operator $\bm{\Upsilon}$:
\begin{equation} \label{qo}
(-\frac{\omega^2}{c^2}+\bm{\Upsilon})\bm{\mathcal Q({\bm{r}},\omega)}=\frac{4\pi \omega^2}{c^2\sqrt{\epsilon({\bm{r}})}}{\bm{P}}\,.
\end{equation}

We are interested in modeling the linear response of the system to an input field ${\bm{\mathcal E_0}}$ in the following way \cite{SakodaPRA2006}:

\begin{equation}\label{final_result}
\left|\frac{{\bm{\mathcal E}}\left({\bm{r}},\omega\right)}{{\bm{\mathcal E_0}}}\right|^2
={\bm{\mathcal F}}\left({\bm{r}},\omega\right){\bm{\mathcal S}}\left(\omega\right)\,,
\end{equation}
with ${\bm{\mathcal S}}\left(\omega\right)$ the emission spectrum and ${\bm{\mathcal F}}\left({\bm{r}},\omega\right)$ a form factor that gives the spatial and spectral dependence of the field emitted outside the cavity. One central quantity in our approach is the Green's function corresponding to the Maxwell equation \eqref{qo}, defined as:
\begin{equation} \label{green}
\bigg[\frac{\omega^2}{c^2}-\bm{\Upsilon}({\bm{r}})\bigg] {\bm{\mathcal G}({\bm{r}},{\bm{r}}^\prime,\omega)}=\delta({\bm{r}}-{\bm{r}}^\prime)\,.
\end{equation}

Inserting Eq. \eqref{green} into Eq. \eqref{Max}, it is straightforward to prove that the following integral equation holds:
\begin{eqnarray}\label{field}
&&{\bm{\mathcal Q}}\left({\bm{r}},\omega\right)=
{\bm{\mathcal Q_0}}\left({\bm{r}},\omega\right)
\\ \nonumber
\\ \nonumber
&&+4\pi \frac{\omega^2}{c^2} \int\int{{d{\bm{r}}^{\prime }d{\bm{r}}^{\prime \prime}{\bm{\mathcal G}({\bm{r}},{\bm{r}}^\prime,\omega)}\frac{\hat{\bm{\chi}}_{QD}\left({\bm{r}}^{\prime },{\bm{r}}^{\prime \prime},\omega\right)}{\sqrt{\epsilon({\bm{r}}^{\prime })}}
{\bm{\mathcal Q}}\left({\bm{r}}^{\prime \prime},\omega\right)}}\nonumber\,,
\end{eqnarray}

where ${\bm{\mathcal Q_0}}$ is the solution  for the bare cavity (i.e. the PHC nanocavity \emph{in the absence of} an embedded QD). When the QD is embedded in the vacuum or in a planar microcavity, an analytical expression for the Green's function holds \cite{Parascandolo2005}. In a PHC nanocavity this is in general not the case. Yet, a very effective simplification is provided by the use of the resolvent representation of the Green's function, in terms of a set of orthogonal eigenmodes $\Phi_n({\bm{r}})$ of the cavity \cite{Economou}:

\begin{equation} \label{GreeDev}
{\bm{\mathcal G}({\bm{r}},{\bm{r}}^\prime,\omega)} = \int{du\frac{\Phi_u({\bm{r}})\Phi^*_u({\bm{r}}^\prime)}{\frac{\omega_u^2}{c^2}-\frac{\omega^2}{c^2}}}\,,
\end{equation}
Equation \eqref{GreeDev} is formally exact. The continuum of eigenmodes reflects the losses of the PHC cavity. The resonant cavity modes arise as sharp resonances in the energy-dependent density of these eigenmodes. A very effective approximation, close to the resonant mode we are interested in, consists in writing explicitly its contribution to (\ref{GreeDev}) and accounting for its finite lifetime through a damping constant $\kappa$. Then, for $\omega \approx \omega_c$ we can write
\begin{eqnarray}
{\bm{\mathcal G}({\bm{r}},{\bm{r}}^\prime,\omega)} \approx \frac{\Phi_0({\bm{r}})\Phi^*_0({\bm{r}}^{\prime })c^2}{2\omega_c(\omega_c-\omega-i\kappa)}+c({\bm{r}},{\bm{r}}^{\prime },\omega)\,.
\end{eqnarray}
Here, $c({\bm{r}},{\bm{r}}^{\prime })$ represents the contribution of all other modes, and is supposed to be small at $\omega \approx \omega_c$. A complete numerical calculation of cavity eigenmodes \cite{Andreani2004}, would allow to test this assumption. Alternatively, our approach can be justified in terms of the \emph{quasi-mode} theory \cite{Scully1990}, by assuming weak coupling between an ideal undamped cavity mode and the vacuum electromagnetic field outside the cavity \cite{Scully1990}. Then, the spectrum can be obtained from Eq. \eqref{field} and an appropriate description of the QD susceptibility
\begin{equation} \label{Chi}
\hat{\bm{\chi}}_{QD}\left({\bf r},{\bf r}^\prime,\omega\right)=
\frac{\mu_{cv}^2}{\hbar}
\Psi\left({\bf r},{\bf r}\right)
\Psi^*\left({\bf r^\prime},{\bf r^\prime}\right)\hat{\bm{\chi}}\left(\omega\right)\,,
\end{equation}
where $\mu_{cv}$ is the Bloch part of the interband dipole matrix element and $\Psi({\bf r}_e,{\bf r}_h)$ is
the electron-hole-pair envelope wave function in the QD.

\section{CQED limit}

The CQED limit describes a two level atom in a high finesse cavity \cite{Carmichael}. This result can be transposed to the case of a solid state oscillator and a microcavity \cite{Andreani1999}. We start from a single electron-heavy-hole transition at $\omega=\omega_0$, with an additional non-radiative linewidth $\gamma_0$ of the bare QD:
\begin{equation} \label{chisingle}
\hat{\bm{\chi}}\left(\omega\right)=\frac{1}{\omega_0-\omega-i\gamma_0}\,.
\end{equation}
We assume that the cavity mode is much more extended than the electron-hole pair wave function, that are thus approximated by dirac-deltas (${\bm{r_0}}$ is the position of the QD). We finally get

{\scriptsize
\begin{eqnarray}\label{spectrum}
&&\left|{\bm{\mathcal E}}\left({\bm{r_0}},\omega\right)\right|^2
\\ \nonumber
\\ \nonumber
&&
=\left|\frac{\omega^2 \Phi_0({\bm{r_0}})}{2\omega_c}
\frac{\Delta \epsilon_c(\omega_0-\omega-i\gamma_0)}{(\tilde{\omega_0}-\omega-i\gamma_{QD})(\omega_c-\omega-i\kappa)-g^2}
\int_{C} {\bm{\mathcal E_B}}\left({\bm{r}}^{\prime }\right)\Phi^*_0({\bm{r}}^{\prime })\right|^2\,,
\end{eqnarray}}
with $\Delta \epsilon_c=\epsilon_M-1$ -- where $\epsilon_M$ is the dielectric constant of the medium in which the cavity is formed, and ${\bm{\mathcal E_B}}$ is an external field entering the system. We have also expressed $c({\bm{r}},{\bm{r}}^{\prime })$ in terms of its real and imaginary parts as $c({\bm{r}},{\bm{r}}^{\prime })=a({\bm{r}},{\bm{r}}^{\prime })+ib({\bm{r}},{\bm{r}}^{\prime })$ and
\begin{equation}\label{rabi}
 \left\{
    \begin{array}{ll}
        \tilde{\omega_0} & =\omega_0-\frac{4\pi\mu_{cv}^2\omega_c^2a({\bm{r_0}},{\bm{r_0}})}{\hbar\epsilon_M c^2} \approx \omega_0
        \ \\
        \\ \gamma_{QD} &=\gamma_0+\gamma_r \\
        \ \\
        g^2&=\frac{2\pi\mu_{cv}^2\omega_c |\Phi_0({\bm{r_0}})|^2}{\hbar\epsilon_M}
    \end{array}
\right.
\end{equation}
with $\gamma_r=\frac{4\pi\mu_{cv}^2\omega_c^2 b({\bm{r_0}},{\bm{r_0}})}{\hbar\epsilon_M c^2}$ the cavity induced damping rate. The quantity $c({\bm{r}},{\bm{r}}^{\prime })$ models the coupling of the QD to all other field modes which, in addition to a small energy shift, are responsible for the intrinsic radiative lifetime of the two-level system, denoted by $\gamma_r$ in the CQED formalism. In the expression for $g^2$, the mode volume enters through the localized cavity mode wave function $\Phi_0({\bm{r_0}})$.

With the assumption of a dipolar field being emitted close to the dot (\emph{virtual oscillating dipole method}, see Ref. \cite{SakodaPRA2006}), i.e.
\begin{equation*}
{\bm{\mathcal E}_B}\left({\bm{r_0}},\omega\right)={\bm{\mathcal E_0}}f({\bm{r}}-{\bm{r_0}}) \frac{1}{\pi}\frac{\gamma}{(\omega-\omega_0)^2+\gamma^2}\,,
\end{equation*}
we get the following result:

\begin{figure}[]%
  \includegraphics*[width=.5\textwidth]{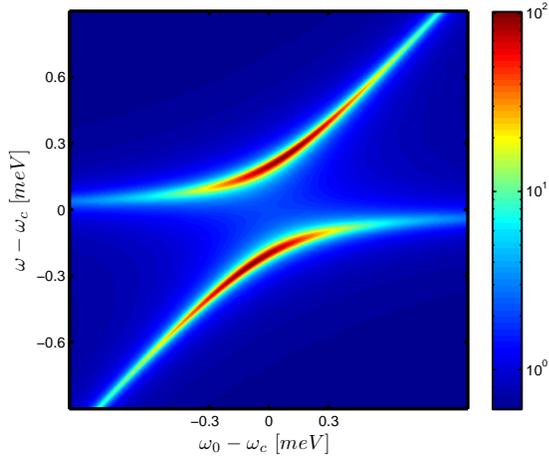}%
  \caption{%
   Response of the system to a dipolar excitation, as given by Eq. \ref{final_result} (logarithmic scale). In this case $\omega_c=1062 \ meV$.}
    \label{fig2}
\end{figure}

\begin{equation}\label{formf}
{\bm{\mathcal F}}\left({\bm{r_0}},\omega\right)={\bm{\mathcal C}}\left|\frac{\omega_c \Phi_0({\bm{r_0}})}{2}\Delta \epsilon_c
\right|^2\,,
\end{equation}

\begin{equation}
{\bm{\mathcal C}}=\left| \int d\bm{r}\Phi_0^*({\bm{r}})f(\bm{r}-\bm{r_0}) \right| ^2\,,
\end{equation}

and
\begin{equation}\label{somega}
{\bm{\mathcal S}}\left(\omega\right)=\left|\frac{\Omega_+-\omega_a+i\gamma}{\omega-\Omega_+}-\frac{\Omega_--\omega_a+i\gamma}{\omega-\Omega_-}\right|^2\,.
\end{equation}
\begin{equation}\label{somega}
\Omega_\pm=\omega_0-\frac{i}{4}(\gamma+\kappa) _\pm \sqrt{g^2-\left(\frac{\gamma-\kappa}{4}\right)^2}\,.
\end{equation}

This is exactly the CQED result \cite{Carmichael,Andreani1999}. As seen in Eq. \eqref{formf}, the form factor ${\bm{\mathcal F}}$ is proportional to the overlap between the QD volume and the cavity mode wave function through ${\bm{\mathcal C}}$. Figure \ref{fig2} displays the response function of the system for different dot-cavity detuning $D=\omega_c-\omega_0$, showing the typical avoided level crossing. As can be seen in Eq. \eqref{rabi}, our model provides a very good quantitative account of the measured Rabi splitting in realistic situations. By using a Gauss-shaped cavity eigenmode, we recover the result of Ref. \cite{Andreani1999}

\begin{equation}\label{somega}
g^2=\frac{2\pi \mu_{cv}^2 \omega_c}{\hbar\epsilon_M V_m}
\end{equation}

\section{Influence of electron-phonon coupling}

In a semiconductor, the QD electronic states are coupled to other degrees of freedom of the surrounding semiconductor. In the limit of low excitation and low temperature, coupling to acoustic phonons dominates. They can be accounted for by using a different expression for $\hat{\bm{\chi}}\left(\omega\right)$.

 \begin{figure}[htb]%
  \includegraphics*[height=.35\textwidth]{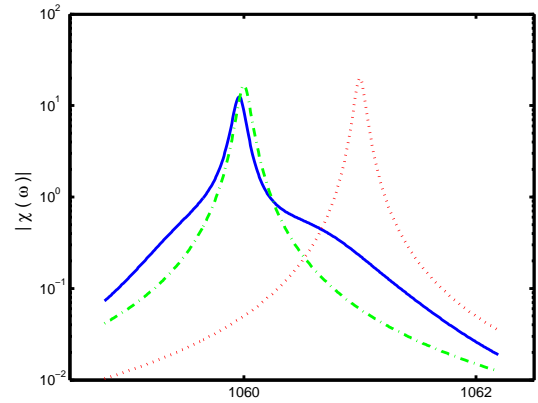}
  \caption{Plot of the quantum dot susceptibility in the presence of phonons and without phonons (dashed line), T=10K. The cavity mode is also visible (dotted line, for $\omega_c=1061 \ meV$). The scale is logarithmic.}
    \label{fig3}
\end{figure}

\begin{figure}[]%
  \includegraphics*[height=.35\textwidth]{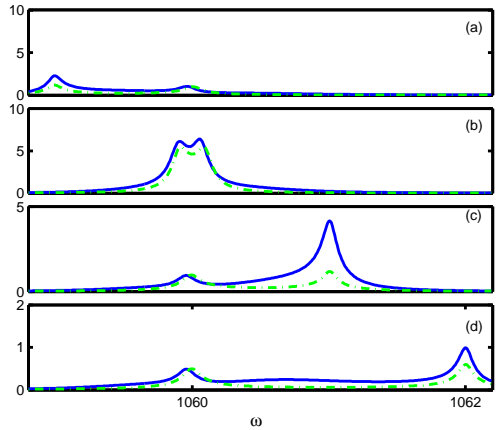}
  \caption{%
   Persistence of the cavity-like peak in the presence of phonons (blue line) for different detunings a) D=-1 meV, b) D=0 meV, c) D=1 meV, d) D=2 meV (T=10K). The result without phonons is plotted as a dashed line ($\omega_c$ is varying).}
    \label{fig4}
\end{figure}

\begin{figure}[]%
  \includegraphics*[height=.4\textwidth]{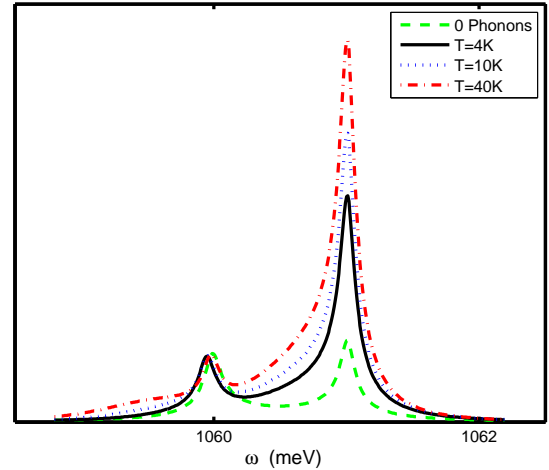}
  \caption{%
  Influence of temperature on the phonon induced effect (arbitrary units, linear scale), $\omega_c=1061 \ meV$.}
    \label{fig5}
\end{figure}

We consider coupling to longitudinal acoustic (LA) phonons via the deformation potential mechanism. This process has already been characterized both theoretically \cite{Zimmermann,Krummheuer} and experimentally \cite{BorriPRL}. Within the second Born approximation and restricting to only one phonon band, the QD linear susceptibility is given by \cite{Krummheuer}:
\begin{equation} \label{Chiph}
\hat{\bm{\chi}}_{QD}^\varphi\left(\omega\right)=\frac{1}{\omega_0-\omega-i\gamma_0+\Sigma(\omega)}\,,
\end{equation}
with
\begin{equation*}
\Sigma(\omega)=\sum_{q}\left[
\frac{{\mid g_{q}^x\mid}^2(1+n(q))}{\omega+i\gamma-\omega_0-\omega(q)}
+
\frac{{\mid g_{q}^x \mid}^2(n(q))}{\omega+i\gamma-\omega_0+\omega(q)}\right]\,,
\end{equation*}

where $g_{q}^x$ is the exciton-phonon coupling matrix element. The absorption of the QD (i.e. the imaginary part of the QD susceptibility) is plotted in fig. \ref{fig3}.
We can observe that phonons contribute with a broad background extending over a few meV, as compared to the bare QD. In the absence of a resonant nanocavity, these sidebands are a vanishingly small feature, difficult to observe in a PL spectrum \cite{Besombes2001}, and requiring non linear coherent spectroscopy to be characterized \cite{BorriPRL}. In a cavity instead, the highly selective spectral filtering of the resonant mode can make this feature dominant. In figure \ref{fig4}, we plot the emission spectrum of the system for different QD-cavity detunings $D$ and with the presence/absence (full line/dashed line) of phonons. For zero detuning (panel (b)) both situation show a Rabi splitting: in this case, the zero-phonon part of the QD susceptibility dominates, and no relevant difference appears. For $D=1 meV$, on the other hand, the cavity-like peak is strongly enhanced in presence of phonon sidebands. This enhancement disappears at still larger detuning, exceeding the width of the phonon sidebands (see panel d)).
This proves that phonons can greatly enhance the emission of the QD at nonzero detuning. One of the main parameters entering this phenomenon is the temperature $T$. In figure \ref{fig5}, we have investigated its influence for a given $D=1 \ meV$ detuning. As expected, the phonon-induced features grow with temperature.

\section{Conclusion}
Many interesting aspects of CQED applied to semiconductor systems have been investigated in the last decade. However, the scenario in which a QD is treated as a \emph{macroatom} is often oversimplified. In this work, we have given a simple example of deviations from the usual CQED results, when the microscopic nature of the electronic states of the QD is taken into account. We have shown that the influence of phonons strongly affects the emission properties of a QD embedded in a PHC nanocavity. Other semiconductor-related mechanisms, like for example Coulomb interaction with the continuum of electronic states present in the wetting layer, would lead to qualitatively similar effects. All these effects should be thoroughly considered in any study of a QD-cavity system aimed at an implementation of quantum information science.

\begin{acknowledgement}
We are grateful to Claudio Andreani and Antonio Badolato for enlightening discussions. We acknowledge
financial support from the Swiss National Foundation through Project No. 200021-117919/1.
\end{acknowledgement}

%
%

\providecommand{\etal}{~et~al.}
\providecommand{\jr}[1]{#1}

\end{document}